\documentclass{ifacconf}

\usepackage{graphicx}      
\usepackage{natbib}        

\usepackage{amsmath} 
\usepackage{amssymb}  

\newtheorem{definition}{Definition}
\newtheorem{remark}{Remark}
\DeclareMathOperator{\sign}{\mathtt{sign}}

\begin{document}
\begin{frontmatter}

\title{Input-to-state Stability of Impulsive Systems with Different Jump Maps\thanksref{footnoteinfo}}

\thanks[footnoteinfo]{This work was supported by the German Federal Ministry
 of Education and Research (BMBF) as a part of the research project ''LadeRamProdukt''.}

\author[First]{Sergey Dashkovskiy}
\author[Second]{Petro Feketa}

\address[First]{University of W{\"u}rzburg,
   W{\"u}rzburg, Germany \\ (e-mail: sergey.dashkovskiy@mathematik.uni-wuerzburg.de)}
\address[Second]{University of Applied Sciences Erfurt,
   Erfurt, Germany \\ (e-mail: petro.feketa@fh-erfurt.de)}

\begin{abstract}                
The paper introduces sufficient conditions for input-to-state stability (ISS) of a class of impulsive systems with jump maps that depend on time. Such systems can naturally represent an interconnection of several impulsive systems with different impulse time sequences. Using a concept of ISS-Lyapunov function for subsystems a small-gain type theorem equipped with a new dwell-time condition to verify ISS of an interconnection has been proven.
\end{abstract}


\end{frontmatter}

\section{Introduction}

Impulsive systems describe processes that combine continuous and discontinuous behavior. A variety of examples can be found in mathematical modelling of applications in logistics, robotics, population dynamics, etc. A basic mathematical theory of impulsive systems as well as fundamental results on existence and stability of solutions can be found in~\cite{SP87} and \cite{LBS89} and references therein.

Nearly at the same time the concept of input-to-state stability (ISS) for systems of ordinary differential equations has been introduced by \cite{Son89}. ISS characterizes a behavior of solutions with respect to external inputs. Later it was also studied for discrete-time systems by \cite{JW01}, switched systems by \cite{MG01} and hybrid systems by \cite{CT05}.

ISS properties of impulsive systems were firstly studied by \cite{HLT05}. Authors provided a set of Lyapunov-based sufficient conditions for establishing ISS with respect to suitable classes of impulsive time sequences. The same approach was used in \cite{HLT08} to justify integral ISS property of impulsive system.

An important question of ISS theory is establishing sufficient conditions for ISS of interconnected systems. The first results on the ISS property were given for two coupled continuous systems by \cite{JTP94} and for an arbitrarily large number ($n\in \mathbb N$) of coupled continuous systems by \cite{DRW07}. Lyapunov versions of these so-called ISS small-gain theorems were proved by \cite{JMW96} (two systems) and by \cite{DRW10} ($n$ systems). Small-gain theorems for impulsive systems with and without time-delay were established by \cite{DKMN12}. A complementary result for infinite-dimensional impulsive systems were developed by \cite{DM13}.

The latest developments in the area of interconnections and systems of a large scale were made in a strongly related class of hybrid systems. We refer the reader to papers of \cite{LNT14, MYL14, San14} with the most recent small-gain theorems on ISS of hybrid systems. However, solutions to hybrid systems are defined on hybrid time domains, as opposed to the usual time defined on the real line. As it was mentioned in \cite{HLT08}, this leads to a distinct notion of ISS, and some systems that are ISS in impulsive framework are not ISS in the hybrid framework. This motivates an importance of investigation of explicitly impulsive systems apart from hybrid ones.

A significant lack of the previously developed results on interconnections of impulsive systems is that impulsive time sequences of each subsystem must coincide in order to apply known results. The reason for this is trivial: a composition of two impulsive systems with different impulsive time sequences falls out of a class of impulsive systems. Such interconnection requires a jump map to be different depending on time. Moreover, even simple real-processes (for instance in logistics, see Example 1) require for a jump map that varies from time to time.

The aim of this paper is to extend the notion of ISS to impulsive systems with jump maps that depend on time. The novelty of our results is twofold. First, such approach enables modelling of a wider class of processes with continuous and discontinuous dynamics. Second, an interconnection of two impulsive systems will become an impulsive system again. This enables a comprehensive stability analysis of interconnections which subsystems have different impulse time sequences.

The rest of the paper is organized as follows. Section 2 contains a short motivation for the framework extension. An extension of impulsive systems to the case of multiple jump sets is proposed in Section 3. A Lyapunov-like sufficient conditions for ISS of impulsive system along with an appropriate dwell-time condition are presented in Section 4. A small-gain type theorem for a network of two impulsive systems with different impulse time sequences is presented in Section 5. A discussion and a short conclusion complete the paper.

\section{Motivation}

Well-known ISS results (\cite{HLT08};\linebreak \cite{DKMN12}; \cite{DM13}) for impulsive systems were developed for a system of the type
\begin{equation}\label{impteel}
\begin{array}{ll}
\begin{cases}
\dot x(t) = f(x(t),u(t)),\\
x(t)=g(x^-(t), u^-(t)),
\end{cases} &
\begin{aligned}
t&\not=t_k, k\in\{1,2,\ldots\}, \\
t&=t_k, k\in\{1,2,\ldots\}
\end{aligned}
\end{array}
\end{equation}
where $\{t_1,t_2,t_3,\ldots\}$ is a strictly increasing sequence of impulse times in $(t_0,\infty)$ for some initial time $t_0$; the state $x(t)\in\mathbb R^N$ is absolutely continuous between impulses; input $u(t)\in\mathbb R^M$ is locally bounded and Lebesgue-measurable function; $f$ and $g$ are functions from $\mathbb R^N \times \mathbb R^M$ to $\mathbb R^N$, with $f$ locally Lipschitz. The state $x$ and the input $u$ are assumed to be right-continuous, and to have left limits at all times. We denote by $(\cdot)^-$ the left-limit operator, i.e., $x^-(t)=\lim_{s\nearrow t}{x(s)}$. However system \eqref{impteel} does not cover a variety of many real processes. Consider the following
\subsubsection{Example 1.}
\begin{figure}[hb]
  \includegraphics[width=0.5\textwidth]{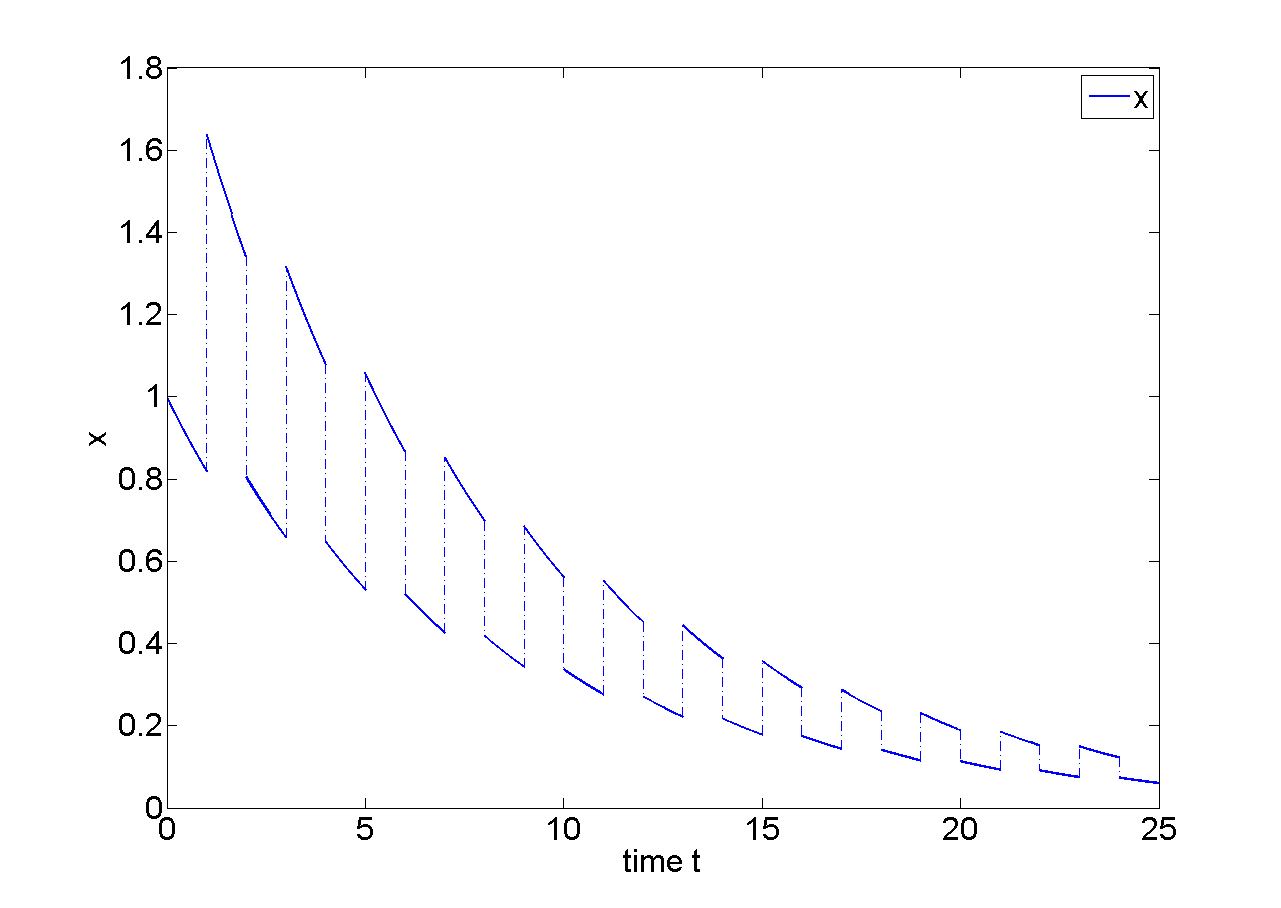}
\caption{Numerical simulation of solution to system \eqref{ex1} with initial condition $x(0)=1$.}
\label{fig1}
\end{figure}

Let the number $x\in\mathbb R_{\geq 0}$ of goods in a storage be continuously decreasing proportionally to the number of items with rate coefficient $0.2$. But every odd day ($\mathbb T_1=\{1,3,5,\ldots\}$) a delivery truck doubles the number of items, and every even day ($\mathbb T_2=\{2,4,6,\ldots\}$) a delivery truck takes out $40\%$ of items. The evolution of this process can be modelled as
\begin{equation}\label{ex1}
\begin{array}{ll}
\begin{cases}
\dot x(t) = -0.2x(t), \\
x(t)=
\begin{cases}
2x^-(t), \\
0.6x^-(t),
\end{cases}
\end{cases} &
\begin{aligned}
t&\not\in\mathbb T_1 \cup \mathbb T_2, \\
t&\in\mathbb T_1, \\
t&\in\mathbb T_2,
\end{aligned}
\end{array}
\end{equation}
System \eqref{ex1} does not fit into the class of systems \eqref{impteel} because its jump map depends on time. Numerical simulations show that a trivial solution to this system is likely to be globally asymptotically stable (see Figure \ref{fig1}).
Moreover, if one adds inputs to the right-hand side of the equations the resulting system becomes ISS, however currently there are no theoretical result to verify this. \hfill$\square$

Another motivation to consider impulsive systems with jump map that depends on time appears while interconnecting several impulsive systems of a type \eqref{impteel} but with different impulse time sequences. Consider two systems of the type \eqref{impteel} with the states $x_i$, inputs $u_i$, flow and jump maps $f_i$ and $g_i$ respectively and impulse time sequences $\mathbb T_i=\{t^i_1,t^i_2,t^i_3,\ldots\}$, $i=1,2$. Let us interconnect these two systems in the following way
\begin{equation*}
u_1=h_2(u_2),\quad u_2=h_1(u_1)
\end{equation*}
with some functions $h_1$ and $h_2$. Then the entire interconnection will have the form
\begin{equation}\label{interconnect}
\begin{array}{ll}
\begin{cases}
\dot x(t) = f(x(t)),\\
x(t)=
\begin{cases}
\tilde g_1(x^-(t)),\\
\tilde g_2(x^-(t)),\\
\tilde g_3(x^-(t)),
\end{cases}
\end{cases} &
\begin{aligned}
t&\not\in\mathbb T_1 \cup \mathbb T_2, \\
t&\in\mathbb T_1, t\not\in\mathbb T_2, \\
t&\not\in\mathbb T_1, t\in\mathbb T_2, \\
t&\in\mathbb T_1, t\in\mathbb T_2,
\end{aligned}
\end{array}
\end{equation}
where $x(t)=(x_1(t),x_2(t))^T$ is the new state, 
$f(x(t))=\left(f_1(x_1(t),h_2(x_2(t))),f_2(x_2(t),h_1(x_1(t)))\right)^T$ is the flow map and functions
\begin{equation*}
\begin{split}
\tilde g_1(x^-(t))&=\left(
\begin{array}{c}
g_1(x_1^-(t), h_2(x_2^-(t)))\\
x_2^-\\
\end{array}
\right),
\\
\tilde g_2(x^-(t))&=\left(
\begin{array}{c}
x_1^-\\
g_2(x_2^-(t), h_1(x_1^-(t)))\\
\end{array}
\right),
\\
\tilde g_3(x^-(t))&=\left(
\begin{array}{c}
g_1(x_1^-(t), h_2(x_2^-(t)))\\
g_2(x_2^-(t), h_1(x_1^-(t)))\\
\end{array}
\right)
\end{split}
\end{equation*}
 define jump maps in different situations (with respect to impulse time sequence). From \eqref{interconnect} it is clear that interconnection of two impulsive systems is a system with three different jump maps depending on impulse time sequence. This simple interconnection cannot be written in the form of \eqref{impteel}.

In the following sections we will develop sufficient conditions of ISS for systems of types \eqref{ex1}, \eqref{interconnect}. This enables a comprehensive stability analysis of a wide class of processes that combine continuous and discontinuous dynamics as well as their interconnections.

\section{Extended Impulsive System}

For every $i=1,\ldots,p$, $p\in\mathbb N$ let $\mathbb T_i = \{t^i_1, t^i_2, t^i_3, \ldots\}$ be a strictly increasing sequence of impulse times in $(t_0,\infty)$ for some initial time $t_0$ and $$\mathbb T = \bigcup_{i=1,\ldots,p}{\mathbb T_i}, \quad \bigcap_{i=1,\ldots,p}{\mathbb T_i}=\emptyset.$$

We consider impulsive system with external inputs
\begin{equation}\label{eis}
\begin{array}{ll}
\begin{cases}
\dot x(t) = f(x(t),u(t)),\\
x(t)=g_i(x^-(t), u^-(t)),
\end{cases}
&
\begin{aligned}
t&\not\in \mathbb T, \\
t&\in \mathbb T_i, i=1,\ldots,p,
\end{aligned}
\end{array}
\end{equation}
where the state $x(t)\in\mathbb R^N$ is absolutely continuous between impulses; input $u(t)\in\mathbb R^M$ is locally bounded and Lebesgue-measurable function; $f$ and $g_i, i=1,\ldots,p$ are functions from $\mathbb R^N \times \mathbb R^M$ to $\mathbb R^N$, with $f$ locally Lipschitz. Given a sequence $\mathbb T_i$, $i=1,\ldots,p$ and a pair of times $s,t$ satisfying $t>s\geq t_0$, let $N_i(t,s)$ denote the number of impulsive times $t^i_k\in\mathbb T_i$ in the semi-open interval $(s,t]$.

As one may see, system \eqref{eis} coincides with the system \eqref{impteel} when $p=1$. Also, similarly to \cite{SP87} and \cite{LBS89},  system \eqref{eis} allows for jump maps $g_i$ to be different for every moment of impulsive jump ($p=\infty$). As it was shown previously, a natural representation of interconnection of two systems of the type \eqref{impteel} requires three sets of impulsive jumps $\mathbb T_1, \mathbb T_2$ and $\mathbb T_3$ with the corresponding jump maps $g_1, g_2$ and $g_3$ ($p=3$).

To introduce appropriate notion of ISS, we recall the following standard definition: a function $\alpha:[0,\infty)\to[0,\infty)$ is of class $\mathcal K$, and we write $\alpha\in\mathcal K$, when $\alpha$ is continuous, strictly increasing, and $\alpha(0)=0$. If $\alpha$ is also unbounded, then we say it is of class $\mathcal K_\infty$, and we write $\alpha\in\mathcal K_\infty$. A continuous function $\beta:[0,\infty)\times[0,\infty)\to[0,\infty)$ is of class $\mathcal{KL}$, and we write $\beta\in\mathcal{KL}$, when $\beta(\cdot,t)$ is of class $\mathcal K$ for each fixed $t\geq 0$, and $\beta(r,t)$ decreases to $0$ as $t\to\infty$ for each fixed $r\geq 0$.

\begin{definition}
For a given set of time sequences $\mathbb T_i$, $i=1,\ldots,p$ of impulse times we call system \eqref{eis} input-to-state stable (ISS) if there exist functions $\beta\in\mathcal{KL}$ and $\gamma\in\mathcal K_\infty$, such that for every initial condition and every input $u$, the corresponding solution to \eqref{eis} exists globally and satisfies
\begin{equation*}
\begin{split}
|x(t)|\leq\max\{\beta(|x(t_0)|, t-t_0),\gamma(\left\|u\right\|_{[t_0,t]})\} \quad \forall t\geq t_0,
\end{split}
\end{equation*}
where $\left\|\cdot\right\|_J$ denotes the supremum norm on an interval $J$.
\end{definition}

\begin{definition}
System \eqref{eis} is called uniformly ISS over a given set $\mathcal S$ of admissible sequences of impulse times if it is ISS for every sequence in $\mathcal S$, with $\beta$ and $\gamma$ independent of the choice of the sequence from the class $\mathcal S$.
\end{definition}

\section{Sufficient Conditions for ISS}

For analysis of ISS of impulsive systems we adopt the concept of a candidate exponential ISS-Lyapunov function from \cite{HLT08} and state it in the implication form.
\begin{definition}
Function $V:\mathbb R^n\to\mathbb R$ is called a candidate exponential ISS-Lyapunov function for \eqref{eis} with rate coefficients $c,d_1,d_2,\ldots,d_p\in\mathbb R$ if $V$ is locally Lipschitz, positive definite, radially unbounded, and whenever $V(x)\geq \gamma(|u|)$ holds it follows that
\begin{align}
\label{expVC} \nabla V(x)\cdot f(x,u)&\leq -cV(x) \qquad \forall x \text{~a.e.}, \forall u,\\
\label{expVD} V(g_i(x,u))&\leq e^{-d_i}V(x) \quad\mbox{~} \forall x,u, i=1,\ldots,p
\end{align}
for some function $\gamma\in\mathcal K_\infty$. In \eqref{expVC}, ''$\forall x \text{~a.e.}$'' should be interpreted as ''for every $x\in\mathbb R^n$ except, possibly, on a set of zero Lebesgue-measure in $\mathbb R^n$''.
\end{definition}

\begin{thm}[Uniform ISS]\label{thm1}
Let $V$ be a candidate exponential ISS-Lyapunov function for \eqref{eis} with rate coefficients $c,d_1,\ldots,d_p\in\mathbb R$ with $d_i\not=0$, $i=1,\ldots,p$. For arbitrary constants $\mu, \lambda>0$, let $\mathcal S[\mu, \lambda]$ denote the class of impulse time sequences satisfying
\begin{equation}\label{dtc}
-\sum_{i=1}^{p}{d_i N_i(t,s)}-(c-\lambda)(t-s)\leq\mu \quad \forall t\geq s\geq t_0.
\end{equation}
Then the system \eqref{eis} is uniformly ISS over $\mathcal S[\mu,\lambda]$.
\end{thm}
\begin{pf}    
The technique of the proof is based on the proof of Theorems 3.3 and 3.7 from \cite{DKMN12}. The difference is that by iteration over $N(t_0,t)$ impulses on interval $(t_0,t]$ we obtain $\sum_{i=1}^p{d_iN_i(t_0,t)}$ instead of $dN(t_0,t)$.
\end{pf}

\begin{remark}
In the case of a distinct jump map for every moment of impulsive jumps (the case of $p=\infty$), dwell-time condition \eqref{dtc} can be presented in the form
\begin{equation*}
-\sum_{t_1^i\in (s,t]}{d_i}-(c-\lambda)(t-s)\leq\mu \quad \forall t\geq s\geq t_0.
\end{equation*}
\end{remark}

\subsubsection{Example 1 (revisited).}
Consider system with external input
\begin{equation}\label{ex1r}
\begin{array}{ll}
\begin{cases}
\dot x(t) = -0.2x(t)+u(t),\\
x(t)=
\begin{cases}
2x^-(t)+u^-(t),\\
0.6x^-(t)+u^-(t),
\end{cases}
\end{cases}&
\begin{aligned}
t&\not\in\mathbb T_1 \cup \mathbb T_2, \\
t&\in\mathbb T_1, \\
t&\in\mathbb T_2
\end{aligned}
\end{array}
\end{equation}
with $\mathbb T_1=\{1,3,5,\ldots\}$ and $\mathbb T_2=\{2,4,6,\ldots\}$. System \eqref{ex1r} possesses candidate exponential ISS-Lyapunov function $V=|x|$ with $c=0.2$, $d_1=-\ln{2}$ and $d_2=-\ln{0.6}$. It satisfies the dwell-time condition \eqref{dtc} with, for instance, $\mu=-d_1=\ln{2}$ and $\lambda=0.1$. So system \eqref{ex1r} is ISS. Moreover, it means that system with zero input \eqref{ex1} is globally asymptotically stable.\hfill$\square$

The dwell-time condition \eqref{dtc} firstly appears in this paper. It generalises the corresponding dwell-time condition from \cite{HLT08} for the case of multiple jump maps. ISS property of system \eqref{ex1r} cannot be verified using previously known results. Using the framework of measure differential equations with discontinuous inputs, a similar sufficient condition for asymptotic stability in terms of vectors fields were derived in~\cite{TBP15}.

\section{Interconnected impulsive systems}

Let two strictly increasing sequences of impulse times $\widehat{\mathbb T}_i = \{t^i_1, t^i_2, t^i_3, \ldots\}$, $i=1,2$ be given and $\mathbb T:=\widehat{\mathbb T}_1 \cup \widehat{\mathbb T}_2$. We consider an interconnection of two impulsive systems with inputs of the form
\begin{equation}\label{intsyst}
\begin{array}{ll}
\begin{cases}
\dot x_i(t) = f_i(x_1(t),x_2(t), u_i(t)),\\
x_i(t)=\hat g_i(x_1^-(t), x_2^-(t), u_i^-(t)),
\end{cases}
\begin{aligned}
t&\not\in \mathbb T, \\
t&\in\widehat{\mathbb T}_i,
\end{aligned}
\end{array}
\end{equation}
$i=1,2$, where the state $x_i(t)\in\mathbb R^{N_i}$ of the $i$th subsystem is absolutely continuous between impulses; $u_i(t)\in\mathbb R^{M_i}$ is a locally bounded Lebesgue-measurable input, and $x_j(t)\in\mathbb R^{N_j}$, $j\not=i$ can be interpreted as internal inputs of the $i$th subsystem. Furthermore, $f_i:\mathbb R^{N_1}\times\mathbb R^{N_2}\times\mathbb R^{M_i}\to \mathbb R^{N_i}$ and $\hat g_i:\mathbb R^{N_1}\times\mathbb R^{N_2}\times\mathbb R^{M_i}\to \mathbb R^{N_i}$, and we assume that the $f_i$ are locally Lipschitz for $i=1,2$. All signals ($x_1,x_2,u_1,u_2$) are assumed to be right continuous and to have left limits at all times.

We define $N:=N_1+N_2$, $M:=M_1+M_2$, $x:=(x_1, x_2)^T$, $u:=(u_1, u_2)^T$, $f:=(f_1, f_2)^T$, and
\begin{equation*}
\begin{array}{ll}
g(x,u)=\begin{cases}
g_1(x,u):=(\hat g_1(x,u_1), x_2)^T,\\
g_2(x,u):=(x_1, \hat g_2(x,u_2))^T,\\
g_3(x,u):=(\hat g_1(x,u_1), \hat g_2(x,u_2))^T,
\end{cases} &
\begin{aligned}
t&\in\mathbb T_1, \\
t&\in\mathbb T_2, \\
t&\in\mathbb T_3
\end{aligned}
\end{array}
\end{equation*}
such that the interconnected system \eqref{intsyst} is of the form \eqref{eis} with $p=3$, $\mathbb T_1:=\widehat{\mathbb T}_1\setminus \widehat{\mathbb T}_2$, $\mathbb T_2:=\widehat{\mathbb T}_2\setminus \widehat{\mathbb T}_1$, and $\mathbb T_3:=\widehat{\mathbb T}_1\cap \widehat{\mathbb T}_2$.

To describe stability properties of the system \eqref{intsyst} we adopt the definitions of ISS and ISS-Lyapunov function for subsystem from \cite{DKMN12}.

\begin{definition}
For given time sequences $\mathbb T_1$ and $\mathbb T_2$ of impulse times we call the $i$th subsystem of \eqref{intsyst} input-to-state stable (ISS) if there exist functions $\beta_i\in\mathcal{KL}$ and $\gamma_{ij}, \gamma_i\in\mathcal K_\infty\cup\{0\}$, such that for every initial condition $x_i(t_0)$ and every input $u_i$, the corresponding solution to \eqref{intsyst} exists globally and satisfies for all $t\geq t_0$
\begin{equation*}
\begin{split}
|x_i(t)|\leq\max\{&\beta_i(|x(t_0)|, t-t_0), \\ &\max_{j,j\not=i}{\gamma_{ij}(\left\|x_j\right\|_{[t_0,t]})}, \gamma_i(\left\|u\right\|_{[t_0,t]})\},
\end{split}
\end{equation*}
where $i,j=1,2$. Functions $\gamma_{ij}$ are called gains.
\end{definition}


Similarly, a Lyapunov function for a system with several inputs is defined as follows. Assume that for each subsystem of the interconnected system \eqref{intsyst} there is a given function $V_i:\mathbb R^{N_i}\to\mathbb R_+$, which is continuous, proper, positive, and locally Lipschitz continuous on $\mathbb R^{N_i}\setminus\{0\}$.
\begin{definition}
For $i=1,2$, the function $V_i$ is called an exponential ISS-Lyapunov function for the $i$th subsystem of \eqref{intsyst} with rate coefficients $c_i, d_i \in\mathbb R$ if, whenever $V_i(x_i)\geq\max\{\max_{j,j\not=i}{\gamma_{ij}(V_j(x_j))}, \gamma_i(|u_i|)\}$ holds, it follows that
\begin{equation}\label{LFC}
\nabla V_i(x_i)\cdot f_i(x,u_i)\leq -c_iV_i(x_i) \quad \forall x \text{~a.e.}, \forall u_i
\end{equation}
and for all $x$ and $u_i$ it holds that
\begin{equation*}
V_i(\hat g_i(x,u_i))\leq \max\{e^{-\hat d_i}V_i(x_i),\max_{j,j\not=i}{\gamma_{ij}(V_j(x_j))}, \gamma_i(|u_i|)\},
\end{equation*}
where $\gamma_{ij}, \gamma_i$ are some functions from $\mathcal K_\infty$.
\end{definition}
In general, even if all subsystems of \eqref{intsyst} are ISS, the whole system may be not ISS. Moreover, since the moments of impulse jumps of each subsystem differ we are unable to employ known small-gain type theorems from \cite{DKMN12} and \cite{DM13}. These theorems consider only one constant $d$ that characterizes the influence of impulses for a given Lyapunov function. In our settings we have a distinct constant for each jump map. The only case to use previously developed results is to majorize system and consider the worst case approach assuming that the state of every subsystem updates at the moments of impulse jumps. This leads to $d:=\min_{i=1,\ldots,p}{\hat d_i}$. However, such approach is too rough, which means that the ISS of interconnected system may not be verified, although the system possesses ISS property. In the following we prove less conservative Theorem~\ref{thm2}. However this theorem also has a limited field of use. It will be discussed in more details in Section 6.

In this paper, to simplify relations and readability, we are concerned with an interconnection of impulsive systems that have exponential ISS-Lyapunov functions $V_i$ with linear gains $\gamma_{ij}(r):=\gamma_{ij}\cdot r$, $\gamma_{ij}, r >0$. However the proposed results can be extended to the case of non-exponential Lyapunov functions with nonlinear gains.

\begin{thm}\label{thm2}
Assume that each subsystem of \eqref{intsyst} has an exponential ISS-Lyapunov function $V_i$, $i=1,2$ with corresponding linear gains $\gamma_{12}$ and $\gamma_{21}$ and rate coefficients $c_1,c_2,\hat d_1\not=0, \hat d_2\not=0$. Let \eqref{intsyst} be written as \eqref{eis}. If
\begin{equation}\label{SGC}
\gamma_{12}\cdot\gamma_{21}<1,
\end{equation}
then there exists a constant $s_1>0$ such that the candidate exponential ISS-Lyapunov function for the whole system \eqref{eis} can be chosen as
\begin{equation}\label{LF}
V(x):=\max\left\{\frac{1}{s_1}V_1(x_1), V_2(x_2)\right\}.
\end{equation}
Its gain is given by $$\gamma(r):=\max\left\{e^{d_1}, e^{d_2}, e^{d_3}, 1\right\}\max\left\{\frac{1}{s_1}\gamma_1(r),\gamma_2(r)\right\},$$ and the rate coefficients are $c:=\min\{c_1,c_2\}$, $d_1:=\min\{\hat d_1,-\ln{\frac{1}{s_1}\gamma_{12}},-\varepsilon\}$, $d_2:=\min\{\hat d_2,-\ln{\gamma_{21}},-\varepsilon\}$, and $d_3:=\min\{\hat d_1,\hat d_2,-\ln{\frac{1}{s_1}\gamma_{12}},-\ln{\gamma_{21}}\}$, where $\varepsilon>0$ is an arbitrary small constant. In particular, for all $\mu,\lambda>0$ such that the dwell-time condition \eqref{dtc} holds, system \eqref{eis} is uniformly ISS over $\mathcal S[\mu,\lambda]$.
\end{thm}
\begin{pf}    
Since small-gain condition \eqref{SGC} is satisfied, it follows from \cite{JMW96} that there exists a constant $\sigma>0$ such that $\gamma_{12}<\sigma<\gamma_{21}^{-1}$. Let us define $V$ as in \eqref{LF} with $s_1:=\frac{1}{\sigma}$ and show that this function is a candidate exponential ISS-Lyapunov function for system \eqref{eis}. It can be easily checked that this function is locally Lipschitz, positive definite, and radially unbounded.

Consider open domains $M_1,M_2\subset\mathbb R^{N}\setminus\{0\}$ defined by
\begin{equation*}
\begin{split}
M_1:=\left\{(x_1,x_2)^T\in\mathbb R^N\setminus \{0\}: \frac{1}{s_1}V_1(x_1)>V_2(x_2)\right\}, \\
M_2:=\left\{(x_1,x_2)^T\in\mathbb R^N\setminus \{0\}: V_2(x_2)>\frac{1}{s_1}V_1(x_1)\right\}.
\end{split}
\end{equation*}
Take an arbitrary $\hat x=(\hat x_1, \hat x_2)^T\in M_1$. It follows that there exists a neighborhood $U$ of $\hat x$ such that $V(x)=\frac{1}{s_1}V_1(x_1)$ is differentiable for almost all $x\in U$. We define $\bar \gamma_1(r):=\frac{1}{s_1}\gamma_1(r)$, $r>0$, and assume that $V(x)\geq \bar \gamma_1(|u|)$. From the small-gain condition \eqref{SGC} it follows that
\begin{equation*}
\begin{split}
V_1(x_1)=s_1V(x)&\geq\max\{\gamma_{12}V(x), s_1\bar\gamma_1(|u|)\} \\
&\geq \max\{\gamma_{12}V_2(x_2),\gamma_1(|u_1|)\}.
\end{split}
\end{equation*}
Then, from \eqref{LFC}, we obtain that for almost all $x$
\begin{equation*}
\dot V(x)=\frac{1}{s_1}\nabla V_1(x_1)\cdot f_1(x,u_1)\leq-\frac{1}{s_1}c_1V_1(x_1)=-c_1V(x).
\end{equation*}
Now take an arbitrary $\hat x\in M_2$. Analogously, there exists a neighborhood $U$ of $\hat x$ such that $V(x)=V_2(x_2)$ is differentiable for almost all $x\in U$. Assume that $V(x)\geq \gamma_2(|u|)$. From the small-gain condition it follows that
\begin{equation*}
\begin{split}
V_2(x_2)=V(x)&\geq\max\{\gamma_{21}s_1V(x), \gamma_2(|u|)\} \\
&\geq \max\{\gamma_{21}V_1(x_1),\gamma_2(|u_2|)\}.
\end{split}
\end{equation*}
Then, from \eqref{LFC}, we obtain that for almost all $x$
\begin{equation*}
\dot V(x)=\nabla V_2(x_2)\cdot f_2(x,u_2)\leq-c_2V_2(x_2)=-c_2V(x).
\end{equation*}
We have shown that for $c=\min\{c_1,c_2\}$ the function $V$ satisfies \eqref{expVC} with $\bar\gamma(r)=\max\{\frac{1}{s_1}\gamma_1(r), \gamma_2(r)\}$, $r>0$ for all $\hat x\in M_1 \cup M_2$.

To treat the points $\hat x\in\mathbb R^N\setminus\{M_1\cup M_2\}$ we need to consider three different cases corresponding to the jump maps $g_1$, $g_2$, and $g_3$:
\begin{equation*}
\begin{split}
&V(g_1(x,u))=V\left( \begin{pmatrix}\hat g_1(x,u_1)\\x_2\end{pmatrix} \right) \\
&=\max\left\{\frac{1}{s_1}V_1(\hat g_1(x,u_1)), V_2(x_2)\right\} \\
&\leq\max\left\{\frac{1}{s_1}\max\{e^{-\hat d_1}V_1(x_1),\gamma_{12}V_2(x_2),\gamma_1(|u_1|)\},V_2(x_2)\right\}\\
&\leq\max\left\{\frac{1}{s_1}\max\{e^{-\hat d_1}s_1V(x),\gamma_{12}V(x),\gamma_1(|u_1|)\},V(x)\right\} \\
&\leq\max\left\{e^{-\hat d_1}V(x),e^{\ln{\frac{1}{s_1}\gamma_{12}}}V(x), e^0V(x),\frac{1}{s_1}\gamma_1(|u_1|)\right\} \\
&\leq\max\{e^{-d_1}V(x), \bar \gamma_1(|u|)\},
\end{split}
\end{equation*}
where $d_1:=\min\{\hat d_1,-\ln{\frac{1}{s_1}\gamma_{12}},-\varepsilon\}$ with an arbitrary small constant $\varepsilon>0$.

Define $\tilde\gamma_1(r):=e^{d_1}\bar\gamma_1(r)$. If it holds that $V(x)\geq\tilde\gamma_1(r)$, it follows that
\begin{equation*}
\begin{split}
V(g_1(x,u))&\leq\max\{e^{d_1}V(x), \bar \gamma_1(|u|)\} \\
&=\max\{e^{d_1}V(x), e^{d_1}\tilde \gamma_1(|u|)\}\leq e^{d_1}V(x)
\end{split}
\end{equation*}
for all $x$ and $u$, and $V$ satisfies \eqref{expVD} with $d_1$ and $\tilde\gamma_1$.

Now consider $V(g_2(x,u))$. In the analogues way one can prove that if it holds that $V(x)\geq\tilde\gamma_2(|u|)$, it follows that
\begin{equation*}
V(g_2(x,u))\leq e^{d_2}V(x)
\end{equation*}
for all $x$ and $u$, and $V$ satisfies \eqref{expVD} with $d_2:=\min\{\hat d_2,-\ln{\gamma_{21}},-\varepsilon\}$ and $\tilde\gamma_2(r):=e^{d_2}\gamma_2(r)$.

Finally, if it holds that $V(x)\geq\tilde\gamma_3(|u|)$, it follows that
\begin{equation*}
V(g_3(x,u))\leq e^{d_3}V(x),
\end{equation*}
for all $x$ and $u$, and $V$ satisfies \eqref{expVD} with $$d_3:=\min\left\{\hat d_1,\hat d_2,-\ln{\frac{1}{s_1}\gamma_{12}},-\ln{\gamma_{21}}\right\}$$ and $\tilde\gamma_3(r):=e^{d_3}\max\{\bar\gamma(r),\gamma_2(r)\}$.

Define $\gamma(r):=\max\left\{e^{d_1}, e^{d_2}, e^{d_3}, 1\right\}\max\left\{\frac{1}{s_1}\gamma_1(r),\gamma_2(r)\right\}$. Then we conclude that $V$ is a candidate exponential ISS-Lyapunov function with rate coefficients $c, d_1, d_2, d_3$ and gain $\gamma$. Hence, from Theorem~\ref{thm1} it follows that system \eqref{intsyst} is uniformly ISS over $\mathcal S[\mu,\lambda]$ if the dwell-time condition \eqref{dtc} holds.
\end{pf}

Note that if system \eqref{intsyst} does not posses external inputs $u_i$, $i=1,2$ then Theorem~\ref{thm2} provides sufficient conditions for global asymptotic stability.

\subsubsection{Example 2.}
Consider the interconnection of two impulsive systems
\begin{equation}\label{ex2}
\begin{cases}
&\begin{aligned}
\dot x(t) &= -(1+\delta)x(1+e^{(1+\delta)|x|})+|y|e^{|y|}\\
\dot y(t) &= -y(1+e^{|y|})+|x|e^{|x|}
\end{aligned}, \mbox{~} t\not=k,\\
&x(t)=3x^-(t), \hfill\quad\qquad\qquad\qquad\qquad\qquad t=2k-1, \\
&y(t)=2y^-(t), \quad\qquad\qquad\qquad\qquad\qquad t=2k
\end{cases}
\end{equation}
where $k\in\mathbb N$, $\delta>0$, the state $(x,y)\in\mathbb R\times\mathbb R$. Consider functions $V_1(x)=|x|$ and $V_2(y)=|y|$. Then
\begin{equation*}
\begin{split}
\dot V_1(x)&=\sign{x}\left(-(1+\delta)x\left(1+e^{(1+\delta)|x|}\right)+|y|e^{|y|}\right) \\
&\leq -(1+\delta)|x|\left(1+e^{(1+\delta)|x|}\right)+|y|e^{|y|} \\
&\leq -(1+\delta)V_1(x)\left(1+e^{(1+\delta)V_1(x)}\right)+V_2(y)e^{V_2(y)} \\
&\leq -(1+\delta)V_1(x)
\end{split}
\end{equation*}
if only
\begin{equation}\label{eq20}
-(1+\delta)V_1(x)e^{(1+\delta)V_1(x)}+V_2(y)e^{V_2(y)}\leq 0.
\end{equation}
Consider function $\xi(s)=se^s\in\mathcal K_{\infty}$. Since for any constant $\alpha\in\mathbb R_+$: $\xi(\alpha s)=\alpha s e^{\alpha s}$ from \eqref{eq20} we get $$\xi\left((1+\delta)V_1(x)\right)\geq\xi\left(V_2(y)\right).$$ Since $\xi\in\mathcal K_\infty$ $\dot V_1(x)\leq -(1+\delta)V_1(x)$ if $V_1(x)\geq\frac{1}{1+\delta}V_2(y)$. We have proven that for the first subsystem there exists an exponential ISS-Lyapunov function with rate coefficient $c_1=1+\delta$ and linear gain function with coefficient $\gamma_{12}=\frac{1}{1+\delta}$. Similarly, one can show that $V_2(y)=|y|$ is an exponential ISS-Lyapunov function for the second subsystem with rate coefficient $c_2=1$ and linear gain function with coefficient $\gamma_{21}=1$. It is obvious that the small-gain condition $\gamma_{12}\gamma_{21}=\frac{1}{1+\delta}<1$ holds. Hence the continuous part of system \eqref{ex2} has an ISS dynamics and the corresponding candidate exponential ISS-Lyapunov function has the rate coefficient $c:=\min\{c_1,c_2\}=1$. For the discrete part, the corresponding rate coefficients of the exponential ISS-Lyapunov functions for subsystems are $\hat d_1=-\ln{3}<0$ and $\hat d_2=-\ln{2}<0$. So the discrete dynamics play against stability of the network. Constant $s_1$ from \eqref{LF} can be chosen in such a way that the rate coefficients of the candidate exponential ISS-Lyapunov function for the entire interconnection are $d_1:=\hat d_1=-\ln{3}$ and $d_2:=\hat d_2=-\ln{2}$. Since the moments of impulsive perturbations do not coincide, from Theorem~ \ref{thm2} system \eqref{ex2} is GAS if there exist constants $\mu,\lambda>0$ such that the dwell-time condition
\begin{equation}\label{dtc_ex}
\ln{3}N_1(t,s)+\ln{2}N_2(t,s)-(1-\lambda)(t-s)\leq\mu
\end{equation}
holds for all $t\geq s\ \geq t_0$. It is easy to verify that there exists a sufficiently small constant $\lambda>0$ such that the dwell-time condition \eqref{dtc_ex} holds for $\mu=\ln{3}$. So system \eqref{ex2} is GAS.\hfill$\square$

Note that previously known theorems are not applicable to the system \eqref{ex2}. There is the only possibility to majorize system \eqref{ex2} assuming $y(t)=3y^-(t)$, so the jump maps of subsystems coincide. Let us denote this system by (\ref{ex2}') and show that this approach is unsuccessful for verifying GAS.  From \cite{DKMN12}, the majorant system (\ref{ex2}') is GAS if
\begin{equation}\label{dtc_ex2}
\ln{3}\left(N_1(t,s)+N_2(t,s)\right)-(1-\lambda)(t-s)\leq\mu
\end{equation}
holds for some $\mu,\lambda>0$. However even for an arbitrary small $\lambda>0$ there does not exist $\mu$ to satisfy \eqref{dtc_ex2} for all $t\geq s\ \geq t_0$. Consider semi-open intervals $(s,t]:=(k,k+2i]$, $i\in\mathbb N$ and the corresponding sequence of $\mu_i$-th that satisfy \eqref{dtc_ex2}:
\begin{equation*}
\mu_i\geq\ln{3}+2i\ln{3}-2i(1-\lambda)=\ln{3}+2i\left(\ln{3}-1+\lambda\right).
\end{equation*}
Since for any $\lambda>0$ the expression $\ln{3}-1+\lambda$ is positive, the sequence $\{\mu_i\}$ is strictly increasing and unbounded. So \eqref{dtc_ex2} does not hold and we cannot verify GAS of the majorant system (\ref{ex2}') and consequently of the original system \eqref{ex2}.

\section{Discussion and Conclusion}
Theorem~\ref{thm2} firstly proposes a small-gain type approach for stability analysis of interconnection of two impulsive systems with different impulse time sequences. A previously known theorems (\cite{DKMN12, DM13}) cannot be directly applied to such interconnection. In some cases a majorizing system with more frequent impulsive jumps of the state of each subsystem can be constructed. Additionally, these jumps should be of the worst nature towards stability: the resulting rate coefficient of the corresponding candidate exponential ISS-Lyapunov function will be estimated as $d\leq\min_{i=1,\ldots,p}{\hat d_i}$. This leads to an excessive conservativeness of the resulting conditions.

In this paper we have developed a more flexible tool for the study of such systems. It is worth to mention that in the case of two different time sequences, the resulting rate coefficient $d_1$ does not depend explicitly on $\hat d_2$ and $d_2$ does not depend explicitly on $\hat d_1$. In some particular cases, new theorem provides fairly accurate sufficient conditions for ISS of the interconnection. It is the case of ISS continuous dynamics ($c_1,c_2>0$) and ''bad'' discrete ones ($\hat d_1, \hat d_2<0$). We would like also to mention that if the times of impulse jumps of subsystems coincide, Theorem~\ref{thm2} reduces to the previously known propositions.

The main weakness of Theorem~\ref{thm2} is that it ignores a positive contribution of impulsive jumps towards ISS of the system. For example, let $\hat d_1>0$. In the case of $t\in\mathbb T_1 \setminus \mathbb T_2$ the second part of the state remains unchanged since it is updated according to the rule $x_2(t)=x_2^-(t)$. Using the worst-case approach, it leads to a conclusion that this impulse does not contribute to stability. That is why the conditions of the Theorem~\ref{thm2} for the case of $d_i$-s of different signs are more conservative, but it is still valid.  Theorem~4.2 from \cite{DKMN12} and Theorem~4 from \cite{MYL14} face a similar problem even to a greater extent since the resulting rate coefficient $d\leq\min_{i=1,\ldots,p}{\hat d_i}$ in theirs setting. It means that even a single jump map with ''bad'' dynamics ($\hat d_i<0$) leads to the ignorance of positive contribution of all other jump maps.

The ignorance of contribution of ''good'' impulses ($\hat d_1, \hat d_2>0$) makes Theorem~\ref{thm2} inapplicable to study the case of unstable continuous dynamics of some subsystem ($c_i<0$) in most cases (excluding completely ''good'' discrete dynamics ($\hat d_1,\hat d_2>0$) and coincidence of impulsive jumps of subsystems), due to inability to satisfy dwell-time condition \eqref{dtc}. One of the potential way to overcome this limitation is to construct a new exponential ISS-Lyapunov function for the subsystems with new rate coefficients $\tilde c_i, \tilde d_i$ such that $\tilde c_i>0$ for all $i$ as it is suggested in \cite{MYL14}. However such approach leads to substantial increase of internal gains. It is out of the scope of current paper, but it is a good way to prolong this research.

The question of how to account and estimate different types of influences caused by different parts of system in a flexible and precise manner still remains open.

\begin{ack}
This work was supported by the German Federal Ministry of Education and Research (BMBF) as a part of the research project ''LadeRamProdukt''.
\begin{figure}[hb]
  \includegraphics[width=1.5in]{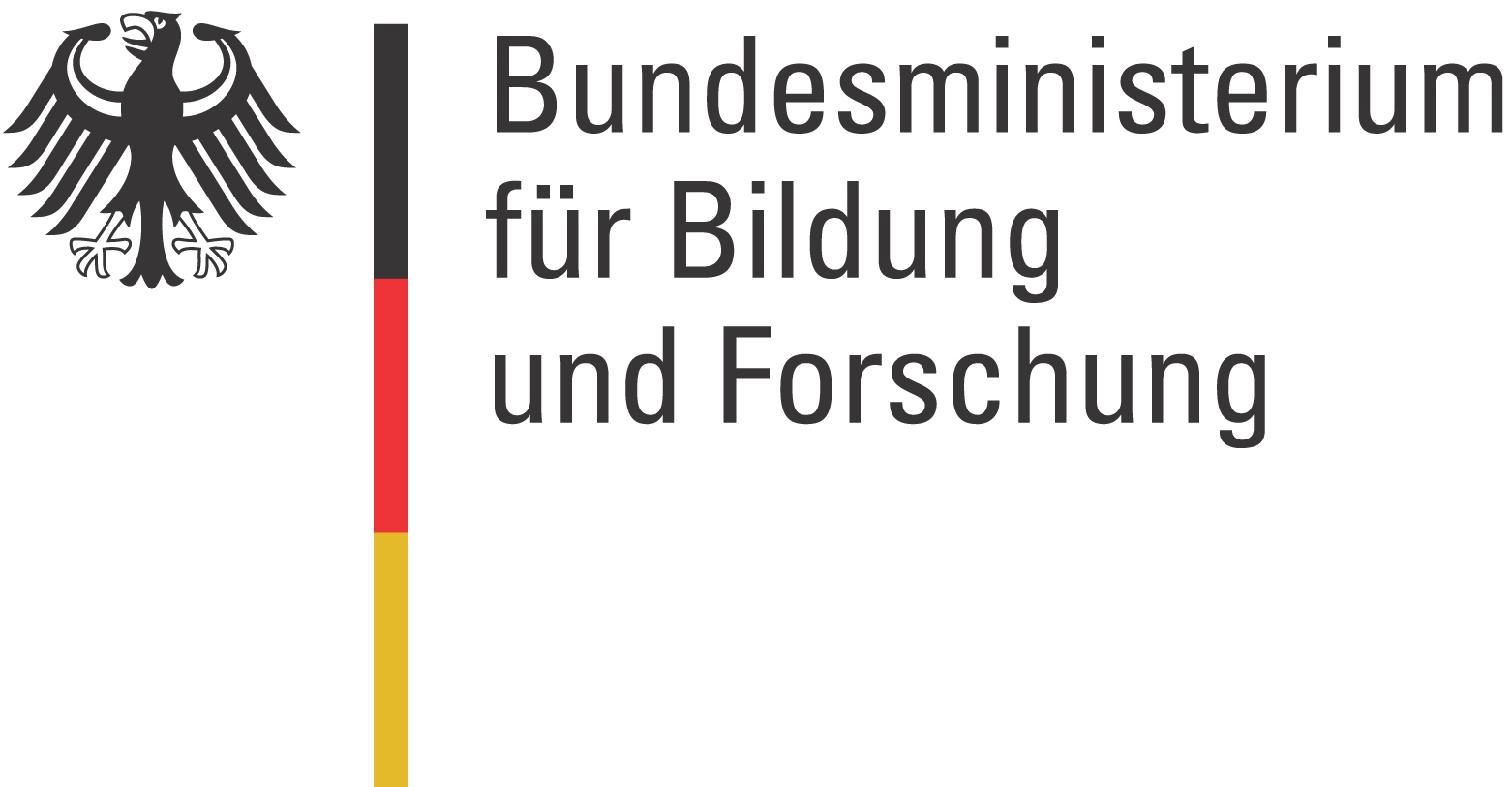}
  \hfill
  \includegraphics[width=1.5in]{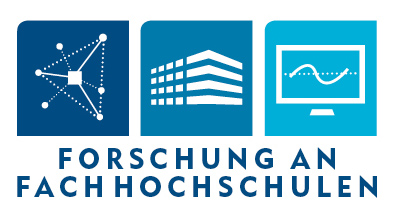}
\end{figure}
\end{ack}


\bibliography{ifacconf}             

\end{document}